\documentclass[12pt]{article}
\usepackage{epsfig,amssymb,amsmath,psfrag}

\def \BES  {(0)}
\def \g  {\chi}

\def \be  {\begin{equation}}
\def \ee  {\end{equation}}
\def \ba  {\begin{eqnarray}}
\def \ea  {\end{eqnarray}}
\def \baa {\begin{eqnarray*}}
\def \eaa {\end{eqnarray*}}
\def \bb  {\begin {thebibliography} }
\def \eb  {\end{thebibliography}}
\def \lab #1 {\label{#1}}

\newcommand\re[1]{(\ref{#1})}
\def \qqquad {\qquad\quad}
\def \qqqquad {\qquad\qquad}
\def \matrix #1 {\left(\begin{array}{cc} #1 \end{array}\right)}

\def \tr {\mathop{\rm tr}\nolimits}

\def \Re {\mathop{\rm Re}\nolimits}

\def \e  {\mathop{\rm e}\nolimits}
\newcommand\lr[1]{{\left({#1}\right)}}

\def\XXint#1#2#3{{\setbox0=\hbox{$#1{#2#3}{\int}$}
     \vcenter{\hbox{$#2#3$}}\kern-.5\wd0}}

\begin{document}

\renewcommand{\thefootnote}{\fnsymbol{footnote}}

\begin{titlepage}
\begin{flushright}
\begin{tabular}{l}
LPT--Orsay--08--44 \\
\end{tabular}
\end{flushright}

\vskip3cm

\begin{center}
 {\large \bf Embedding nonlinear $\rm O(6)$ sigma model \\[2mm] into $\mathcal{N}=4$ super-Yang-Mills theory}
\end{center}

\vspace{1cm}

\centerline{\sc B.~Basso, G.P.~Korchemsky}

\vspace{10mm}

\centerline{\it Laboratoire de Physique Th\'eorique\footnote{Unit\'e
                    Mixte de Recherche du CNRS (UMR 8627).},
                    Universit\'e de Paris XI}
\centerline{\it 91405 Orsay C\'edex, France}

\vspace{1cm}

\centerline{\bf Abstract}

\vspace{5mm} Anomalous dimensions of high-twist Wilson operators have a nontrivial scaling behavior
in the limit when their Lorentz spin grows exponentially with the twist. To describe the
corresponding scaling function in planar $\mathcal{N}=4$ SYM theory, we analyze an integral equation
recently proposed by Freyhult, Rej and Staudacher  and argue that at strong coupling it can be
casted into a form identical to the thermodynamical Bethe Ansatz equations for the nonlinear $\rm
O(6)$ sigma model. This result is in a perfect agreement with the proposal put forward by Alday and
Maldacena within the dual string description, that the scaling function should coincide with the
energy density of the nonlinear $\rm O(6)$ sigma model embedded into $\rm AdS_5\times S^5$.

\end{titlepage}

\setcounter{footnote} 0


\newpage

\renewcommand{\thefootnote}{\arabic{footnote}}

\section{Introduction}

Anomalous dimensions of high-twist Wilson operators have a nontrivial scaling behavior in the limit
when their Lorentz spin grows exponentially with the twist~\cite{BGK06}. Recently, Alday and
Maldacena~\cite{AM07} used a dual stringy description of such operators in $\mathcal{N}=4$
super-Yang-Mills theory (SYM) to put forward the proposal that, at strong coupling,  the
corresponding scaling function should coincide in a suitable limit with the energy density of a
two-dimensional bosonic $\rm O(6)$ sigma-model. In this paper, we establish the same relation on the
gauge theory side of the AdS/CFT correspondence.

The operators under consideration are built from $L$ complex scalar fields $X(0)$ and $N$ covariant
derivatives $D_+=(n\cdot D)$ projected onto the light-cone direction $n_\mu^2=0$
\begin{align}\label{O}
\mathcal{O}_{N,L} (0) = \tr \left[ D_+^{k_1} X(0)  D_+^{k_2} X(0) \ldots  D_+^{k_L} X(0)\right] ,
\end{align}
with $N=k_1+\ldots +k_L$. These operators mix under renormalization and their anomalous dimensions,
defined as eigenvalues of the mixing matrix, have a rich structure. For given $N$ and $L$, they
occupy a band and the properties of anomalous dimensions are different in the lower and upper part
of the spectrum~\cite{BGK03}. In what follows we shall study the {\em minimal anomalous dimension}
$\gamma_{N,L}(g)$ in the planar $\mathcal{N}=4$ SYM theory, both at weak and strong coupling, in the
limit $N, L\to\infty$. The motivation for considering this particular limit is twofold and it goes
beyond the scope of $\mathcal{N}=4$ model. High spin/twist Wilson operators analogous to \re{O}
naturally appear in QCD (with a different partonic content though) in the operator product expansion
description of deeply inelastic scattering in the semi-inclusive regime, the so-called large Bjorken
$x$ limit~\cite{Gardi02}. On the other hand, in planar $\mathcal{N}=4$ SYM, the Wilson operators
\re{O} admit a dual description in terms of folded strings spinning on the $\rm AdS_5\times
S^5$~\cite{GKP,FT} and their anomalous dimensions can be found at strong coupling from the AdS/CFT
correspondence~\cite{Mal97}.

The minimal anomalous dimension $\gamma_{N,L}(g)$  is a complicated function of the Lorentz spin
$N$, twist $L$ and 't Hooft coupling $g^2= {\lambda}/{(4\pi)^2} =  {g_{\rm YM}^2 N_c}/{(4\pi)^2}$.
For finite twist $L$ and large spin $N$ it has a remarkable logarithmic (Sudakov) scaling
behavior~\cite{K89,BGK03}
\be\label{old}
\gamma_{N,L}(g) = 2\Gamma_{\rm cusp}(g)\ln N+O(N^0)\,,\qquad N\to\infty,\ L={\rm fixed}\,,
\ee
where the coupling dependent prefactor does not depend on the twist $L$ and is given by the cusp
anomalous dimension~\cite{P80,K89}. The situation changes however when $N$ and $L$ become large
simultaneously. In that case, the anomalous dimension $\gamma_{N,L}(g)$ still scales logarithmically
for sufficiently large $N$. However, in distinction with \re{old}, the value of the spin for which
transition into the Sudakov regime takes place now depends both on the twist $L$ and on the coupling
constant. As was shown in Ref.~\cite{BGK06}, different regimes of the anomalous dimension
$\gamma_{N,L}(g)$ at large $L$ and $N\gg L$ are controlled by the order parameter $\xi$, which is
given at weak coupling by $\xi(g<1) = \ln({N}/{L})/L$ and at strong coupling by $\xi(g\gg
1)=g\ln({N}/{L})/L$:
\begin{itemize}
\item
For $\xi<1$  the minimal anomalous dimension has a BMN like~\cite{BMN} scaling%
\footnote{As was shown in Ref.~\cite{FTT06}, this scaling breaks down at higher orders in the BMN
coupling $g^2/L^2$.} both at weak and strong coupling~\cite{FT,BFST03,BGK06}
\be\label{BMN}
\gamma_{N,L} = 8g^2 \ln^2(N/L)/L+O(g^4\ln^4(N/L)/L^3)\,,\qquad  {N,L\to\infty \,,~\xi< 1}\,.
\ee
\item
For $\xi\gg 1$  the minimal anomalous dimension grows logarithmically with $N$ and it has the
following scaling behavior both at weak and at strong coupling~\cite{BGK06,FTT06,CK07,AM07,FRS}%
\footnote{Our definition of $\epsilon(g,j)$ and $j$ is slightly different  compared to that of
Ref.~\cite{AM07}.}
\be\label{sum_up}
\gamma_{N,L}(g) = \left[ 2\Gamma_{\rm cusp}(g)+ \epsilon(g,j)\right] \ln N + \ldots\,,\quad \quad
 {N,\,L\to\infty\,,~j = \frac{L}{\ln N}={\rm fixed}}\,.
\ee
\end{itemize}
Here $\epsilon(g,j)$ is a nontrivial function of the scaling variable $j$ satisfying
$\epsilon(g,j=0)=0$ and ellipses denote terms suppressed by powers of $1/L$. We note that the
condition $\xi\gg 1$  automatically implies that  $j \ll 1$ at weak coupling and $j\ll g$ at strong
coupling.

At weak coupling, the scaling function $\epsilon(g,j)$ can be found in a generic (supersymmetric)
Yang-Mills theory in the planar limit by making use of the remarkable property of integrability
\cite{L94,FK95}. In QCD, the dilatation operator in the $SL(2)$ sector is integrable for the special
class of maximal helicity operators to two loops at least~\cite{BDM98,B99,BKM04}, whereas in
$\mathcal{N}=4$ SYM theory it is believed that integrability gets extended in this sector to all
loops~\cite{BS03,S}. The scaling function has the following form at weak coupling
\be\label{weak}
\epsilon(g,j) = \epsilon_1(g) j + \epsilon_2(g) j^2 + \epsilon_3(g) j^3 + \ldots\,,
\ee
with the coefficient functions $\epsilon_k(g)$ given by series in $g^2$. To one-loop order, one
finds in planar $\mathcal{N}=4$ theory~\cite{BGK06} \footnote{We displayed in \re{a13} the
dependence on $s$ to indicate that the same expressions hold in a generic (planar) Yang-Mills theory
in integrable $SL(2)$ sector of aligned helicity operators built from gauge fields ($s=3/2$) and
gaugino ($s=1$).}
\be\label{a13}
\epsilon_1 = 4\left[\psi(s)-\psi(2s)\right] g^2+\ldots \,,\qquad \epsilon_2=0\cdot g^2 +\ldots
\,,\qquad \epsilon_3 = \frac{\pi^2}{24}\lr{-\psi''(s)} g^2+\ldots
\ee
where $s=1/2$ is the conformal spin of scalar fields entering \re{O} and
$\psi(x)=\lr{\ln\Gamma(x)}'$ is the Euler psi-function. Recently, an integral equation has been
proposed by Freyhult, Rej and Staudacher (FRS) \cite{FRS} to describe the function $\epsilon(g,j)$
for arbitrary values of the scaling parameter $j$ and the coupling constant $g$. It generalizes the
BES equation~\cite{BES} which governs the dependence of the cusp anomalous dimension $\Gamma_{\rm
cusp}(g)$ on the coupling constant in planar $\mathcal{N}=4$ SYM.  One of the consequences of this
equation is that $\epsilon_2(g)=0$ to any order in $g^2$.

At strong coupling, the gauge/string correspondence relates the minimal anomalous dimension to the
energy of a folded string spinning on the $\rm AdS_5\times S^5$ background~\cite{GKP,FT}
\be
\Delta=N+L+\gamma_{N,L}(g)\,,
\ee
with $N$ and $L$ being angular momenta on $\rm AdS_3$ and $\rm S^1$, respectively. Semiclassical
quantization of this state yields the expansion of the scaling function $\epsilon(g,j)$ in powers of
$1/g$. The first two terms of this expansion have been computed in Refs.~\cite{FTT07,CK07,RT07}.
Recently, Alday and Maldacena~\cite{AM07} put forward the proposal that the scaling function
$\epsilon(g,j)$ can be found {\em exactly} at strong coupling in the limit $j\ll g$ and $j/m={\rm
fixed}$ (with the parameter $m$ defined below in \re{m_AM}). They argued that quantum corrections in
the $\rm AdS_5\times S^5$ sigma model are dominated in this limit by the contribution of massless
excitations on $\rm S^5$ whose dynamics is described by a (noncritical)  two-dimensional bosonic
$\rm O(6)$ sigma-model equipped with a UV cut-off determined by the mass of massive excitations.

The $\rm O(6)$ sigma model emerges within the AdS/CFT as an effective two-dimensional theory
describing the scaling function $\epsilon(g,j)$ in the $\rm AdS_5\times S^5$ sigma model. More
precisely, the quantity
\be
\epsilon_{\rm O(6)}  = \frac{\epsilon(g,j) + j}2=\lim_{N\to\infty} \frac{\Delta-N}{2\ln N}  -  \Gamma_{\rm
cusp}(g)
\ee
has the meaning of the energy density in the ground state of the $\rm O(6)$ model corresponding to
the particle density $\rho = L/(2\ln N) = j/2$. The $\rm O(6)$ sigma model is an exactly solvable
theory~\cite{ZZ78}. It has a nontrivial dynamics in the infrared and massless excitations acquire
mass through dimensional transmutation mechanism~\cite{PW83,FR85}. The exact value of the mass gap
was found using the Bethe Ansatz in Ref.~\cite{HMN90}. In terms of parameters of the underlying $\rm
AdS_5\times S^5$ sigma model, its value reads~\cite{AM07}
\be\label{m_AM}
m = k g^{1/4} \e^{-\pi g} \left[ 1 + O(1/g)\right],\qquad k=\frac{2^{3/4} \pi^{1/4}}{\Gamma(5/4)}\,.
\ee
The dependence of $m$ on the coupling constant is fixed by the two-loop  beta-function of the $\rm
O(6)$ model whereas the prefactor $k$ was determined in \cite{AM07} by matching the first few terms
of $1/g$ expansion of $\epsilon_{\rm O(6)}$ into semiclassical expansion of $\epsilon(g,j)$ computed
in~\cite{FTT06}.

The appearance of mass gap in the dual stringy description of the scaling function has dramatic
consequences for $\mathcal{N}=4$ SYM at strong coupling. It suggests that the strong coupling
expansion of the anomalous dimensions in $\mathcal{N}=4$ SYM should receive nonperturbative
corrections characterized by a new `hidden' scale $m$. This is a rather unexpected and surprising
feature given the fact that the anomalous dimensions capture dynamics at short distances and their
weak coupling expansion is free of nonperturbative corrections. Indeed, such nonperturbative
corrections have been identified in Ref.~\cite{BKK07} in the strong coupling expansion of the cusp
anomalous dimension 
\be\label{pert}
\Gamma_{\rm cusp}(g) =  \sum_{k=-1}^\infty c_k/g^{k} + \alpha\, m^2 + o(m^2)\,,
\ee
where the perturbative coefficients $c_k$ grow factorially at large $k$ and  the value of $\alpha$
depends on regularization of Borel singularities in the perturbative series. The $O(m^2)$
corrections are exponentially small at strong coupling but they provide a leading contribution in
the region $g\sim 1$ in which the transition from the strong to weak coupling regime takes place.

In this paper, we solve the FRS equation \cite{FRS} using the approach developed in
Ref.~\cite{BKK07} and evaluate the scaling function $\epsilon(g,j)$ in $\mathcal{N}=4$ SYM theory at
strong coupling. More precisely, we demonstrate that, in the limit $g\to\infty$ and $j/m={\rm
fixed}$, the integral equation for $\epsilon(g,j)$ can be casted into a form identical to the
thermodynamical Bethe Ansatz equations for the nonlinear $\rm O(6)$ sigma model~\cite{HMN90}.  This
allows us to determine $\epsilon(g,j)$ at strong coupling  for different values of the ratio $j/m$
and to translate nontrivial properties of the $\rm O(6)$ sigma model (mass gap generation at large
distances and asymptotic freedom at short distances) into the corresponding scaling behavior of
$\epsilon(g,j)$:
\begin{itemize}

\item For  $j\ll m \ll g$, or equivalently $N \gg \e^{L/m}$,
\be\label{11}
\epsilon(j,g) +j  =  m^2\left[\frac{j}{m} + \frac{\pi^2}{24} \lr{\frac{j}{m}}^3+
O\left(j^4/m^4\right)\right] .
\ee
In this expansion, $O(j)$ terms are in agreement with the numerical solution to the FRS equation
found in \cite{FGR08}. Also, $O(j^2)$ term is absent, in a striking similarity with vanishing of the
function $\epsilon_2(g)$ in the weak coupling expansion \re{weak} and \re{a13}. Note that the
nonperturbative corrections in \re{pert} and \re{11} are of the same order in $m^2$ and are added
together in the expression for the anomalous dimension \re{sum_up}.

\item For $m \ll j\ll g$, or equivalently $ \e^{L/m} \gg N \gg \e^{L/g} $,
\be
\epsilon(j,g)+j = j^2 \left[ \frac{\pi}{8\ln(j/m)}+ O\lr{\frac{\ln\ln(j/m)}{\ln^2(j/m)}}\right].
\ee
This expression resums through renormalization group (an infinite number of) perturbative
corrections in $1/g$ with coefficients proportional to $j^2$ and enhanced by powers of
$\ln(j/g)$~\cite{AM07,RT07}.

\item For $j\ll g$ and $j/m={\rm fixed}$, or equivalently $N\sim  \e^{L/m}$, the function $\epsilon(j,g)$
does not admit a simple representation and it can be found as the solution to thermodynamical Bethe
Ansatz equations for the $\rm O(6)$ model \cite{HMN90} (see Eqs.~\re{ex1} -- \re{map} below).

\end{itemize}
These results are in a perfect agreement with the Alday-Maldacena proposal~\cite{AM07} and,
therefore, constitute a nontrivial test of the AdS/CFT correspondence.

The paper is organized as follows. In Sect.~2, we reformulate the integral equation proposed in
Ref.~\cite{FRS} and argue that it can be significantly simplified by employing a nontrivial change
of variables found in \cite{BKK07}. In Sect.~3 we work out the expansion of the scaling function at
small $j$. We demonstrate that at weak coupling it agrees with the known one-loop result, whereas at
strong coupling it matches the string theory prediction by Alday and Maldacena including expression
for the mass gap \re{m_AM}. In Sect.~4 we show that the integral equation for the scaling function
can be mapped for $g\to\infty$ and $j/m=\rm fixed$ into thermodynamical Bethe Ansatz equations for
the energy density in the ground state of two-dimensional $\rm O(6)$ sigma-model with the particle
density $j/2$. Section 5 contains concluding remarks. Some details of our calculations are
summarized in two appendices.

\section{The scaling function in $\mathcal{N}=4$ SYM}

The derivation of the scaling function relies on integrability of the dilatation operator in the
$SL(2)$ sector. To one-loop order, the function $\epsilon(g,j)$ was determined in Ref.~\cite{BGK06}
from the analysis of Bethe Ansatz equation in the scaling limit $N, L\to\infty$ and $j={\rm fixed}$.
The Bethe Ansatz solution for $\epsilon(g,j)$ is characterized in this limit by two sets of
parameters, the Bethe roots and the so-called small roots of the transfer matrix, which describe
excitations dubbed `magnons' and `holes', respectively. As was shown in \cite{BGK06}, both sets of
parameters form a dense distribution on the real axis with the holes confined to the interval
$[-a,a]$ and magnons to the union of two intervals $[-\infty,-a]\cup [a,\infty]$. The scaling
function is uniquely determined by the corresponding distribution densities of holes and magnons.

This analysis was recently extended to all loops in \cite{FRS} by employing the asymptotic Bethe
Ansatz approach proposed in Refs.~\cite{AFS04,S}. It led to the FRS equation
\be\label{FRS}
\hat\sigma(t) = \frac{t}{\e^t-1}\lr{\hat{\mathcal{K}}(t,0)-4\int_0^\infty dt' \,
\hat{\mathcal{K}}(t,t')\, \hat\sigma(t')}\,,
\ee
whose solution  $\hat\sigma(t)=\hat\sigma(t;g,j)$ is related to the scaling function as
\be 
\label{f=sigma} f(g,j) \equiv 2\Gamma_{\rm cusp}(g) + \epsilon(g,j)=  j+ 16 \hat\sigma(0)\,.
\ee
The kernel $\hat{\mathcal{K}}(t,t')$ is given by a rather complicated expression that can be found
in Ref.~\cite{FRS} (see also Eq.~\re{K-hat} below). It takes into account a nontrivial scattering
phase which satisfies the crossing symmetry~\cite{J06} and whose explicit form was found in
\cite{BT05}. Later in the paper we shall use another, equivalent formulation of the integral
equation \re{FRS} which is more suitable for studying the strong coupling limit.

For $j\to 0$ the integral equation \re{FRS} reduces to the BES equation \cite{BES} and its solution
$\hat\sigma(t;g,j=0)$ determines the cusp anomalous dimension $f(g,j=0) = 2\Gamma_{\rm cusp}(g)$.
Then, the scaling function $\epsilon(g,j)$ admits the representation
\be\label{ff}
\epsilon(g,j) = f(g,j) - f(g,0)\,.
\ee
At weak coupling the function $f(g,j)$ admits an expansion in powers of $g^2$ and it vanishes as
$g^2\to 0$. Then, it follows from \re{f=sigma}  that the integral equation \re{FRS} has a nontrivial
solution at $g=0$ satisfying $\hat\sigma(0;g=0,j) =-j/16$.

\subsection{Integral equation}

It is convenient to split $\hat\sigma(t)$ into a sum of three terms
\begin{align}\label{sigma}
\hat\sigma(t) = \frac1{\e^t-1}\left[ \e^{-t/2} \gamma_{\rm h}(t)+\frac{g}2\gamma(2gt)-\frac{j}{8}
J_0(2gt)\right]\,,
\end{align}
with $J_0(x)$ being the Bessel function, and rewrite \re{FRS} as a system of coupled integral
equations for the functions $\gamma_{\rm h}(t)$ and $\gamma(2gt)$
\begin{align} \label{gamma_h}
\gamma_{\rm h}(t) & = K_{\rm h}(t,0) - 4\int_0^\infty dt'\, K_{\rm h}(t,t')\e^{t'/2}\,\hat\sigma(t')\,,
\\ \notag
{\gamma(2gt)} &= \frac{2t}{g}\left[ K(t,0) - 4\int_0^\infty dt'\, K(t,t')\,\hat\sigma(t')\right],
\end{align}
with the kernels $K_{\rm h}$ and $K$ defined below in Eqs.~\re{K_h} and \re{K_h1}, respectively. The
rationale behind the decomposition \re{sigma} is that, as we will show below, the functions
$\gamma_{\rm h}(t)$ and $\gamma(2gt)$ have different analytical properties. Namely,  the Fourier
transforms of $\gamma_{\rm h}(t)$ and  $\gamma(2gt)$ defined as solutions to \re{gamma_h} have the
support on the intervals $[-a,a]$  and $[-2g,2g]$, respectively, with the parameter $a$ depending on
$g$ and $j$.

The first relation in \re{gamma_h} involves the kernel
\begin{align} \label{K_h}
K_{\rm h}(t,t') & = \frac{t\cos{(at')}\sin{(at)}-t'\cos{(at)}\sin{(at')}}{2\pi(t^2-t'^2)}
\\ \nonumber
& = \frac1{4\pi} \left[\frac{\sin(a(t-t'))}{t-t'} +\frac{\sin(a(t+t'))}{t+t'}  \right],
\end{align}
and the kernel in the second relation in \re{gamma_h} is defined as
\begin{align} \label{K_h1}
K(t,t') &= g^2 K^{\BES}(2gt,2gt') - 4g^2 \int_0^\infty dt''\,  K^{\BES}(2gt,2gt'') K_{\rm
h}(t'',t')\e^{(t'-t'')/2},
\\ \notag
K^{\BES}(t,t') & = K_+(t,t') + K_-(t,t') + 8g^2 \int_0^\infty \frac{dt''\,t''}{\e^{t''}-1}\,
K_-(t,2gt'') K_+(2gt'',t')\,,
\end{align}
where $K^{\BES}(t,t')$ coincides with the BES kernel \cite{BES} and parity even/odd kernels $K_\pm
(-t,t') = K_\pm (t,-t')= \pm K_\pm(t,t')$ are given by
\begin{align}\label{Kpm}
& K_+(t,t') =\frac{t J_1(t) J_0(t')-t' J_0(t) J_1(t') }{t^2-t'^2}=\frac2{tt'}\sum_{n\ge 1} (2n-1)
J_{2n-1}(t) J_{2n-1}(t')\,,
\\ \notag
& K_-(t,t') =\frac{t' J_1(t) J_0(t')-t J_0(t) J_1(t') }{t^2-t'^2}=\frac2{tt'}\sum_{n\ge 1} (2n)
J_{2n}(t) J_{2n}(t')\,.
\end{align}
Finally, the parameter $a=a(g,j)$ is related to the solution of the integral equation as
\be\label{a(j)}
j = \frac{4a}{\pi} -\frac{16}{\pi}\int_0^\infty dt\,\hat\sigma(t) \e^{t/2}\frac{\sin(at)}{t}\,.
\ee
Being combined together, the relations \re{sigma} -- \re{a(j)} are equivalent to the integral
equation \re{FRS} with the kernel given by
\be\label{K-hat}
\hat{\mathcal{K}}(t,t') = K(t,t') + t^{-1} K_{\rm h}(t,t') \e^{(t'-t)/2} -\frac{J_{0}(2gt)}{t}
\frac{\sin{(at')}}{2\pi t'} e^{t'/2} \,.
\ee
We would like to stress that the functions $\gamma(t)$ and $\gamma_{\rm h}(t)$ depend on the
coupling constant $g$ and the scaling variable $j$. The relations \re{gamma_h} -- \re{a(j)} simplify
significantly as $j\to 0$. In this limit, we have $a\sim j$ so that the kernel $K_{\rm h}(t,t')$ and
the function $\gamma_{\rm h}(t)$ vanish simultaneously as $j\to 0$ leading to
$\hat{\mathcal{K}}_{j=0}(t,t') =g^2 K^{\BES}(2gt,2gt')$ and to the BES equation for the function
$\gamma_{j=0}(2gt)$.

To determine the scaling function \re{f=sigma}, we have to solve the integral equations \re{gamma_h}
and, then, substitute the solution for $\hat\sigma(t)$, Eq.~\re{sigma}, into \re{f=sigma}. Let us
apply  \re{sigma}  to evaluate $\hat\sigma(0)$. We put $t=0$ in both sides of the first relation in
\re{gamma_h} and take into account  the relations \re{K_h} and \re{a(j)} to verify that $\gamma_{\rm
h}(0) = j/8$ for arbitrary $j$ and $g$. Substituting this relation into \re{sigma} and \re{f=sigma}
we find the following representation for the scaling function
\be\label{f-lim}
f(g,j) = 16 g^2 \lim_{t\to 0} \frac{\gamma(2gt)}{2gt}\,.
\ee

\subsection{Parity decomposition}

Similar to the BES equation~\cite{KL07,Alday07}, the integral equations \re{gamma_h} can be further
simplified by separating expressions on both sides of \re{gamma_h} into terms with a definite parity
under $t\to -t$.

According to definition \re{K_h}, the kernel $K_{\rm h}(t,t')$ is an even function of $t$ and $t'$.
Then, it immediately follows from the first relation in \re{gamma_h} that  $\gamma_{\rm h}(t)$ is
an even function of $t$
\be\label{gh-even}
\gamma_{\rm h}(-t) = \gamma_{\rm h}(t)\,,\qquad \gamma_{\rm h}(0) = j/8 \,,
\ee
where the second relation holds for arbitrary coupling $g$. As follows from \re{K_h1}, the kernel
$K(t,t')$ does not have a definite parity under $t\to -t$ and, as a consequence, the function
$\gamma(t)$ can be decomposed as
\be\label{g-dec}
\gamma(t) = \gamma_+(t) + \gamma_-(t)\,,\qquad \gamma_\pm(-t) = \pm \gamma_\pm(t)\,.
\ee
We substitute \re{g-dec} into the second relation in \re{gamma_h} and match parity even/odd terms in
both sides of \re{gamma_h} to obtain after some algebra the system of coupled equations for the
functions $\gamma_\pm(t)$%
\footnote{In comparison with the second relation in \re{gamma_h}, these equations do not involve the
convolution of the kernels $K^{(0)}$ and $K_{\rm h}$ which appears in the kernel $K(t,t')$,
Eq.~\re{K_h1}.}
\be\label{FRS1}
\frac{\gamma_\mp (2gt)}{2g t} =  -4 \int_{0}^{\infty}dt' \, K_{\pm}(2gt, 2gt')\left[
\e^{-{t'}/{2}}\gamma_{\rm h}(t')+\hat{\sigma}(t')\right]+ b_{\pm}(2gt)\,,
\ee
where the kernels $K_\pm$ are defined in \re{Kpm}, $\hat\sigma(t)$ is given by \re{sigma} and the
notation was introduced for
\be
b_+(t) =K_{+}(t, 0)\,,\qquad b_-(t) =4 g \int_{0}^{\infty}\frac{dt'}{\e^{t'}-1} \, K_{-}(t, 2gt')
 {\gamma_-(2gt')}\,.
\ee
We recall that the kernels $K_\pm$ are given by sum over the product of Bessel functions,
Eq.~\re{Kpm}. Substitution of \re{Kpm} into \re{FRS1} yields the expression for $\gamma_\pm(t)$ in
the form of Neumann series over Bessel functions
\begin{align}\label{Bessel}
\gamma_{-}(t) &=  2 \sum_{n\geqslant 1} \  (2n-1)  J_{2n-1}(t) \gamma_{2n-1} \,,
\\ \notag
\gamma_{+}(t) &=   2\sum_{n\geqslant 1} \  (2n) \ J_{2n}(t) \gamma_{2n} \,,
\end{align}
with the expansion coefficients $\gamma_n=\gamma_n(g,j)$ given by the following expressions
\be\label{g=h}
\gamma_{n} =\frac12 \delta_{n,1}+h_n -   \int_{0}^{\infty}\frac{dt}{t} \ \frac{J_{n}(2gt)}{\e^{t}-1}
\left[ {\gamma_+(2gt)-(-1)^n\gamma_-(2gt)} \right].
\ee
Here the notation was introduced for the coefficients $h_n=h_n(g,j)$
\be\label{h_n}
h_n =-\frac2{g}  \int_{0}^{\infty}\frac{dt}{t} \
\frac{J_{n}(2gt)}{\e^{t}-1} \left[
  \e^{{t}/{2}}  \gamma_{\rm h}(t) -\frac{j}{8}J_0(2gt)  \right].
\ee
We recall that the scaling function \re{f-lim} is determined by the asymptotic behavior of the
function $\gamma(t)$ for $t\to 0$. Substituting \re{g-dec} and \re{Bessel} into \re{f-lim} and
taking into account that $J_n(t) \sim t^n$ at small $t$, we find the scaling functions \re{f=sigma}
and \re{ff}
\be\label{f=g1}
f(g,j) =  16 g^2 \gamma_1(g,j)\,,\qquad \epsilon(g,j)  = 16 g^2
\left[\gamma_1(g,j)-\gamma_1(g,0)\right],
\ee
where $\gamma_1$ is the lowest order coefficient in the series \re{Bessel}, $\gamma_-(t) =
\gamma_1(g,j) t + O(t^2)$.

We can obtain another representation for the expansion coefficients $\gamma_n$ in \re{Bessel} by
making use of orthogonality of the Bessel functions
\be\label{ortho}
(2n-1)\int_0^\infty \frac{dt}{t}\, J_{2n-1}(t) J_{2m-1}(t)= 2n\int_0^\infty \frac{dt}{t}\, J_{2n}(t)
J_{2m}(t) = \frac12\delta_{nm}\,.
\ee
In this way, we obtain from \re{Bessel}
\be\label{inverse}
\gamma_{2n-1} = \int_0^\infty \frac{dt}{t} J_{2n-1}(t) \gamma_-(t) \,,\qquad
\gamma_{2n} = \int_0^\infty \frac{dt}{t} J_{2n}(t) \gamma_+(t) \,.
\ee
Here we have tacitly assumed that the Bessel series on the right-hand side of \re{Bessel} are
convergent on the real axis and that the sum over $n$ can be interchanged with the integral over $t$.%
\footnote{In particular, performing the Fourier transformation on both sides of \re{Bessel} and
using well-known properties of the Bessel functions, we find that the Fourier transforms of the
functions $\gamma_\pm(2gt)$ have support on the interval $[-2g,2g]$.} Matching \re{inverse} into
\re{g=h} we obtain an (infinite) system of equations for the functions $\gamma_\pm(t)$
\begin{align}\label{FRS2}
& \int_{0}^{\infty}\frac{dt}{t} \, J_{2n-1}(2gt) \left[
\frac{\gamma_{-}(2gt)}{1-\e^{-t}}+\frac{\gamma_{+}(2gt)}{\e^{t}-1}\right] = \frac{1}{2} \ \delta_{n,
1} +h_{2n-1}(g,j)\,,
\\ \notag
& \int_{0}^{\infty}\frac{dt}{t} \, J_{2n}(2gt) \left[
\frac{\gamma_{+}(2gt)}{1-\e^{-t}}-\frac{\gamma_{-}(2gt)}{\e^{t}-1}\right] = h_{2n}(g,j)\,,
\end{align}
valid for $n\geqslant 1$. These relations should be supplemented with expressions \re{h_n} for
inhomogeneous terms $h_n$. The latter depend on the function $\gamma_{\rm h}(t)$ which depends on
its turn on $\gamma(t)=\gamma_+(t)+\gamma_-(t)$ and satisfies the integral equation \re{gamma_h}
\be\label{gh-new}
\gamma_{\rm h}(t)   = K_{\rm h}(t,0) - 2\int_0^\infty \frac{dt'\, K_{\rm h}(t,t')}{\sinh(t'/2)}
  \left[ \frac{g}2\gamma(2gt')+\e^{-t'/2} \gamma_{\rm h}(t') -\frac{j}{8} J_0(2gt')\right],
\ee
with the kernel $K_{\rm h}(t,t')$ given by \re{K_h}.

Let us now examine the system of integral equations \re{FRS2} and \re{gh-new} for $j=0$. In this
case, the kernel $K_{\rm h}(t,t')$ vanishes and we deduce from \re{gh-new} and \re{h_n} that
$\gamma_{\rm h}(t;j=0)=h_n(g,j=0)=0$. Then, the relations \re{FRS2} are reduced to the BES
equation~\cite{BES} and we denote their solution as
\be\label{g=cusp}
\gamma_\pm(t;j=0)  \equiv \gamma_\pm^{\BES} (t)=t\,\Gamma_{\textrm{cusp}}(g)/(8g^2)+ O(t^2)\,.
\ee
The functions $\gamma_\pm^{\BES} (t)$ were constructed at weak coupling in Refs.~\cite{BES,B06} and
at strong coupling in Refs.~\cite{Alday07,Kostov07,BKK07,B08,KSV08}.

To evaluate the scaling function \re{f=g1} for arbitrary $j$, it is sufficient to determine the
coefficient $\gamma_1$ defined in \re{Bessel} rather than the function $\gamma_-(t)$ itself. This
can be achieved by using the fact that the solutions to the integral equations \re{FRS2} and
\re{gh-new} satisfy Wronskian-like relations. These relations allow us to obtain the following
relation for the scaling function (see Appendix A for details)
\be\label{epsilon-g}
\epsilon(g,j)   =
 32{g}  \int_{0}^{\infty}\frac{dt}{t} \
\frac{\gamma_+^{\BES}(2gt)-\gamma_-^{\BES}(2gt)}{\e^{t}-1} \left[
  \e^{{t}/{2}}  \gamma_{\rm h}(t;j) -\frac{j}{8}\right]
  -4gj \int_0^\infty \frac{dt}{t} \gamma_+^{\BES}(2gt)  \,.
\ee
It is remarkable that $\epsilon(g,j)$ only depends on the solution of the BES equation,
$\gamma_\pm^{\BES}(2gt)$, and the `hole' function $\gamma_{\rm h}(t,j)$. It is the latter function
that generates a nontrivial dependence of $\epsilon(g,j)$ on the scaling parameter $j$. We note that
the integrals on the right-hand side of \re{epsilon-g} are well-defined for $t\to 0$ in virtue of
$\gamma^{\BES}_-(t)\sim t$, $\gamma^{\BES}_+(t) \sim t^2$ and $\gamma_{\rm h}(0;j) = j/8$,
Eqs.~\re{Bessel} and \re{gh-even}, respectively.

\subsection{Change of variables}

The system of integral equations \re{FRS2} has been analyzed  in Refs.~\cite{BKK07,KSV08} in the
special case $h_n=0$ corresponding to $j=0$. The starting point in \cite{BKK07} was the change of
variables $\gamma_\pm(t) \to \Gamma_\pm(t)$ that eliminated the dependence of the BES kernel on the
coupling constant. It is remarkable that the same change of variables also applies to \re{FRS2} for
arbitrary $j$
\be
\Gamma_\pm(t;j) =\gamma_\pm(t;j) \mp \coth\lr{\frac{t}{4g}} \gamma_\mp(t;j)\,,
\ee
or conversely
\be\label{gG}
2\gamma_\pm(t;j) = \lr{1-\frac1{\cosh(t/2g)}} \Gamma_\pm(t;j) \pm \tanh \lr{\frac{t}{2g}}
\Gamma_\mp(t;j)\,.
\ee
Upon this change of variables, the relations \re{FRS2} take the form  (for $n\ge 1$)
\be\label{G-eq}
\int_0^\infty \frac{dt}{t}J_n(t) \bigg[\Gamma_-(t)+(-1)^n\Gamma_+(t) \bigg] = \delta_{n,1} + 2h_n(g,j)\,,
\ee
where $h_n-$coefficients are given by \re{h_n} and the integral kernel on the left-hand side is the
same as for $j=0$. It is convenient to apply the Jacobi-Anger expansion%
\footnote{More precisely, we differentiate both sides of this relation with respect to $u$ and,
then, divide by $t$.} (with $u=\sin\varphi$ and $\varphi$ real)
\be
\e^{it u}  - 1 = 2\sum_{n\ge 1}\left\{ J_{2n}(t) \left[\cos(2n\varphi)-1\right] + i J_{2n-1}(t) \sin
\lr{(2n-1)\varphi}\right\},
\ee
and replace an infinite system of equations \re{G-eq} into a single relation depending on the
$u-$parameter
\be\label{G-int}
\int_0^\infty dt\, \bigg[ \e^{itu} \Gamma_-(t) - \e^{-itu}\Gamma_+(t)\bigg] = 2 - 8  \int_{0}^{\infty}
\frac{dt\,\e^{2ig tu}}{\e^{t}-1} \left[
  \e^{{t}/{2}}  \gamma_{\rm h}(t) -\frac{j}{8}J_0(2gt)  \right].
\ee
It is important to stress that this relation only holds for $-1 \le u \le 1$. Here $\Gamma_\pm(t)$
and $\gamma_{\rm h}(t)$ are real functions of $t$. Taking real and imaginary part on both sides of
\re{G-int}, we can replace \re{G-int} by a system of two coupled equations. Later in the paper we
shall make use of this system.

For $j=0$ the integral on the right-hand side of \re{G-int} vanishes. Its solution at strong
coupling, $\Gamma_\pm(t;j=0)\equiv \Gamma_\pm^{\BES}(t)$, was constructed in
Refs.~\cite{BKK07,KSV08}. For $j\neq 0$ the relation \re{G-int} should be supplemented with integral
equation \re{gh-new} for the function $\gamma_{\rm h}(t)$.

\section{Scaling function at small $j$}

To find the scaling function \re{epsilon-g}, we have to solve a complicated system of coupled
integral equations for the functions $\gamma_{\rm h}(t)$ and $\gamma_\pm(t)$. In this section we
shall evaluate the scaling function $\epsilon(g,j)$ at small $j$ and generalize consideration to
arbitrary $j$ in the next section. As we will see in a moment, the first few terms of the small $j$
expansion of $\epsilon(g,j)$ can be found exactly without solving the integral equations for the
functions $\gamma_\pm(t)$.

\subsection{Small $j$ expansion}

To begin with let us consider the integral equation \re{gamma_h} for the function $\gamma_{\rm
h}(t)$. It involves the kernel $K_{\rm h}(t,t')$ defined in \re{K_h}. According to its definition,
$K_{\rm h}(t,t')$ depends on the parameter $a$  which depends on its turn on $j$ and satisfies the
relation \re{a(j)}
\be\label{j-sample}
j = \frac{4a}{\pi} -\frac{8}{\pi}\int_0^\infty \frac{dt}{t}\, \frac{\sin(at)}{\sinh(t/2)}\left[
\frac{g}2\gamma(2gt)+\e^{-t/2} \gamma_{\rm h}(t)-\frac{j}{8} J_0(2gt)\right]\,.
\ee
We recall that the parameter $a$ determines the interval $[-a,a]$ on which holes rapidities condense
in the limit $N, L\to\infty$ and $j={\rm fixed}$. As was shown in \cite{BGK06}, for $j\to 0$ this
interval shrinks into a point  $a\sim j$ and the scaling function vanishes $\epsilon(g,j=0)=0$.

Assuming that $a\sim j$ we expand the kernel $K_{\rm h}(t,t')$ in powers of $a$ and obtain from
\re{K_h}
\be
K_{\rm h}(t,t') =\frac1{2\pi}\left[a-\frac16 a^3\lr{t^2+t'^2}+ O(a^5) \right]=
\frac{\sin(at')}{2\pi t'} \left[1-\frac16 (at)^2\right]+ O(a^5)\,,
\ee
where expansion runs in odd powers of $a$ only. We substitute this relation into the right-hand side
of \re{gh-new} and observe that the $t'-$integral there coincides with analogous integral on the
right-hand side of \re{j-sample}. Combining the two relations together, we find from \re{gh-new}
\be\label{gh-small}
\gamma_{\rm h}(t) = \frac{j}{8}\left[ 1- \frac16 (at)^2+ O(a^4)\right],
\ee
in agreement with \re{gh-even}.

To determine $a$ we return to the relation \re{j-sample} and expand its both sides in powers of $a$
and $j$. We notice that the last two terms in the square brackets there vanish linearly as $j\to 0$
while the first term  is given by  $\gamma(2gt)=\gamma^{\BES}(2gt) + O(j)$. In this way,
 we find from \re{j-sample}
\be\label{aa}
a=\frac{j\pi}{2\kappa} + O(j^2)\,,
\ee
where the $g-$dependence resides in the normalization factor
\be\label{kappa}
\kappa = 2 -2g\int_0^\infty dt \, \frac{\gamma_+^{\BES}(2gt)+\gamma_-^{\BES}(2gt)}{\sinh(t/2)}\,.
\ee
Substitution of \re{aa} into \re{gh-small} yields the expansion of $\gamma_{\rm h}(t)$ at small $j$.

According to \re{epsilon-g}, the $j-$dependence of the scaling function   $\epsilon(g,j)$ is
controlled by the function $\gamma_{\rm h}(t)$. Replacing  $\gamma_{\rm h}(t)$ in \re{epsilon-g} by
its small $j$ expansion \re{gh-small}, we get
\be\label{e-expansion}
\epsilon(g,j) = j\, \epsilon_1(g) + j^3\,\epsilon_3(g) + O(j^4)\,,
\ee
where the coefficient in front of $j^2$ equals zero for any $g$ and the coefficient functions
$\epsilon_1(g)$ and $\epsilon_3(g)$ are given by
\begin{align}\label{eps13}
\epsilon_1(g) &= -4g   \int_{0}^{\infty}\frac{dt}{t} \left[
\frac{\gamma_+^{\BES}(2gt)}{\e^{-t/2}+1} +\frac{\gamma_-^{\BES}(2gt)}{\e^{t/2}+1}\right],
\\ \notag
\epsilon_3(g)  & =
 -\frac{\pi^2 g}{12\kappa^2}   \int_{0}^{\infty} {dt}\,{t} \
\frac{\gamma_+^{\BES}(2gt)-\gamma_-^{\BES}(2gt)}{\sinh(t/2)}  \,.
\end{align}
We would like to stress that these relations hold for arbitrary coupling $g$.

\subsection{Weak coupling}

We recall that the functions $\gamma_\pm^{\BES}(t)$ satisfy the BES equation. At weak coupling, this
equation can be solved by iterations leading to~\cite{BES,B06}
\begin{align}\label{g-BES}
\gamma_-^{\BES} (t) &= \left(1-g^2\frac{\pi^2}{3}\right) J_1(t) + {O}\left(g^4\right) \,,
\\[2mm]\notag
\gamma_+^{\BES} (t) &= 4g^3\zeta_3 J_{2}(t)+ {O}\left(g^5\right).
\end{align}
Plugging  these expressions into \re{eps13} and expanding $\epsilon_1(g)$ and $\epsilon_3(g)$ in
powers of $g^2$ we get
\begin{align}\label{2loop}
\epsilon_1(g) &= -8 g^2  \ln 2 + g^4\lr{\frac83 \pi^2\ln 2+16\zeta_3} + O(g^6)\,,
\\ \notag
\epsilon_3(g) &= g^2 \frac{7}{12} \pi^2\zeta_3 +
g^4\left(\frac{35}{36}\pi^4\zeta_3-\frac{31}{2}\pi^2\zeta_5\right) + O(g^6)\,.
\end{align}
It is easy to see that the one-loop corrections to  $\epsilon_1(g)$ and $\epsilon_3(g)$ are in an
agreement with the relation \re{a13} for $s=1/2$.

It is straightforward to evaluate higher order corrections to $\epsilon_1(g)$ and $\epsilon_3(g)$ by
taking into account subleading terms on the right-hand side of \re{g-BES}. It is interesting to
notice that higher order corrections to $\epsilon_1(g)$ proportional to `$\ln 2$' are controlled to
all loops by the cusp anomalous dimension
\be\label{ln2}
\epsilon_{1}(g) = - 2\Gamma_{\textrm{cusp}}(g)  \ln 2 + \ldots,
\ee
Indeed, such term originates from the second integral on the right-hand side of \re{eps13} with
$\gamma_-^{\BES}(2gt)$ replaced by its leading small$-t$ asymptotic behavior \re{g=cusp}.

The relations \re{2loop} and \re{ln2} are in agreement with the results of \cite{FRS}.

\subsection{Strong coupling}

To evaluate $\epsilon_1(g)$ and $\epsilon_3(g)$ at strong coupling, it is convenient to substitute
$\gamma_\pm^{\BES}(t)$ on the right-hand side of \re{eps13} by their expressions \re{gG} in terms of
the functions $\Gamma_\pm^{\BES}(t)\equiv\Gamma_\pm(t;j=0)$.

This leads to the following representation for $\epsilon_1(g)$
\begin{align}\label{epsilon1}
\epsilon_1(g) = -{2g}\int_{0}^{\infty}\frac{dt}{t}
\bigg[\left(1-\frac{\cosh\lr{t/4g}}{\cosh\lr{t/2g}}\right)\Big(\Gamma^{\BES}_{-}(t)+\Gamma^{\BES}_{+}(t)\Big)
+ \frac{\sinh\lr{t/4g}}{\cosh\lr{t/2g}} \Big(\Gamma^{\BES}_{-}(t)-\Gamma^{\BES}_{+}(t)\Big)&\bigg].
\end{align}
The functions $\Gamma_\pm(t;j)$ satisfy the integral equation \re{G-int}. For $j=0$ the integral on
the right-hand side of \re{G-int} vanishes and the resulting equations for  $\Gamma_\pm^{\BES}(t)$
can be written as
\begin{align}\label{G-BES}
\int_0^\infty dt\, \sin(ut) \left[\Gamma^{\BES}_{-}(t)+\Gamma^{\BES}_{+}(t)\right]&=0\,,
\\\notag
\int_0^\infty dt\, \cos(ut) \left[\Gamma^{\BES}_{-}(t)-\Gamma^{\BES}_{+}(t)\right] &=2\,,
\end{align}
where $u-$parameter satisfies the condition $u^2\le 1$.

We observe that the relations \re{epsilon1} involve the same combinations of $\Gamma_\pm^{\BES}(t)$
as in \re{G-BES}. To make use of \re{G-BES}, it is suggestive to replace the two factors on the
right-hand side of \re{epsilon1} involving the ratios of hyperbolic functions by their Fourier
integrals. Going through calculation (the details  can be found in Appendix B) we find
\begin{align}\label{eps1=3terms}
\epsilon_{1}(g) =  - {4\sqrt{2}g^2} \bigg\{  &\int_{-1}^{1}du \ u \ \frac{\sinh\lr{g\pi
u}}{\cosh\lr{2g\pi u}}
\\ \notag
+& \int_{1}^{\infty}du \ \frac{\cosh\lr{g\pi u}}{\cosh\lr{2g\pi u}} \int_{0}^{\infty}\frac{dt}{t}
\ \Big(1-\cos\lr{ut}\Big) \Big[\Gamma^{\BES}_{-}(t)+\Gamma^{\BES}_{+}(t)\Big]
 \\ \notag
+& \int_{1}^{\infty}du \ \frac{\sinh\lr{g\pi u}}{\cosh\lr{2g\pi u}} \int_{0}^{\infty}\frac{dt}{t} \
\sin\lr{ut} \ \Big[\Gamma^{\BES}_{-}(t)-\Gamma^{\BES}_{+}(t)\Big]\bigg\}\,.
\end{align}
Here we split the Fourier integral into sum of two terms corresponding to $0\le u\le 1$ and $1\le u
<\infty$ and applied \re{G-BES} in the first term.

Let us now examine the relation \re{eps1=3terms} at strong coupling. It is straightforward to work
out the asymptotic expansion of the first term on the right-hand side of \re{eps1=3terms} at large
$g$. In the remaining two terms, we replace the ratios of hyperbolic functions by their large $g$
expansion and find after some algebra  (see details in Appendix B)
\begin{align}\label{eps1=m}
\epsilon_{1}(g) =  -1 + m + O\left(\e^{-3g\pi}\right)\,,
\end{align}
where the parameter $m=m(g)$ is defined as
\be\label{m=int}
m =  \frac{8\sqrt{2}}{\pi^2}\e^{-\pi g}-\frac{8g}{\pi}\e^{-\pi g}\Re\left[ \int_0^\infty \frac{dt\,
\e^{i(t-\pi/4)}}{t+i\pi g}\left(\Gamma^{\BES}_{+}(t)+ i \Gamma^{\BES}_{-}(t)\right)\right]\,.
\ee
We recall that the functions $\Gamma^{\BES}_{\pm}(t)$ are solutions to the integral equations
\re{G-BES}.

According to \re{eps1=m}, the function $\epsilon_1(g)$ does not receive perturbative corrections in
$1/g$ and the leading nontrivial correction is given by $m$ which is exponentially small in $g$.
This property has the same origin as the appearance of the gap in the distribution of magnon
rapidities at strong coupling described by the BES equation~\cite{BES,Alday07,KSV08}. Namely, the
distribution density of Bethe roots vanishes on the interval $[-2g,2g]$ to each order in
perturbative $1/g$ expansion. These are the nonperturbative (exponentially small in the coupling)
corrections that generate a nonzero distribution density on this interval.

At large $g$ the integral in \re{m=int} receives a dominant contribution from large $t\sim g$. In
order to evaluate \re{m=int} it suffices to take $t=\pi g \tau$ and to replace the functions
$\Gamma^{\BES}_{\pm}(\pi g \tau)$ by their asymptotic behavior for $g\to\infty$ and $\tau={\rm
fixed}$. The leading asymptotic behavior of $\Gamma^{\BES}_{\pm}(\pi g \tau)$ in this limit
reads~\cite{BKK07,KSV08} \footnote{To obtain this relation from perturbative solution for
$\Gamma^{\BES}_{+}(t)+i\Gamma^{\BES}_{-}(t)=\sum_{k\ge 0} g^{-k} \Gamma_{k}(t) $, one has to take
into account that $\Gamma_{k}(t)\sim t^k \Gamma_0(t)$ at large $t$ and, then, resum an entire series
in the double scaling limit $g, t \to\infty$ and $t/g={\rm fixed}$.}
\be\label{G-as}
\Gamma^{\BES}_{+}(\pi g \tau)+i\Gamma^{\BES}_{-}(\pi g \tau) =
-\frac{\sqrt{2}}{\pi}\frac{\Gamma(3/4)\Gamma(1+i\tau/4)}{\Gamma(3/4+i\tau/4)}\left[\int_{-1}^{1}du
\, \left(\frac{1+u}{1-u}\right)^{1/4} \e^{-i\pi g u \tau} + \ldots \right],
\ee
where ellipses denote terms suppressed by powers of $1/g$. From the point of view of the underlying
BES equation~\cite{KSV08}, this expression reflects nontrivial properties of the scattering phase of
magnons in planar $\mathcal{N}=4$ SYM theory in the the near flat space limit~\cite{MS06}.

We substitute \re{G-as} into \re{m=int}, expand the resulting twofold integral at large $g$ and find
after some algebra (see Appendix B)
\be\label{mass}
m = g^{1/4} \e^{-\pi g} \frac{2^{3/4}\pi^{1/4}}{\Gamma(5/4)}+\ldots
\ee
This relation is in a remarkable agreement with the expression for the mass gap \re{m_AM} found by
Alday and Maldacena \cite{AM07} from string theory considerations. Moreover, the obtained expression
for  $\epsilon_1(g)$, Eq.~\re{eps1=m}, is in agreement with the numerical solution to the FRS
equation and its analytical estimate~\cite{FGR08,Benna}.

Let us now evaluate $\epsilon_3(g)$. We start with the second relation in \re{eps13} and change
variables following \re{gG}
\begin{align} \label{e3}
\epsilon_3(g)  & = \frac{\pi^2}{48 g\kappa^2} \int_{0}^{\infty}dt \, t \left[
\frac{\sinh\lr{t/4g}}{\cosh\lr{t/2g}}
\Big(\Gamma^{\BES}_{-}(t)-\Gamma^{\BES}_{+}(t)\Big)-\frac{\cosh\lr{t/4g}}{\cosh\lr{t/2g}}
\Big(\Gamma^{\BES}_{-}(t)+\Gamma^{\BES}_{+}(t)\Big)\right],
\end{align}
and analogously for $\kappa$ defined  in \re{kappa}
\be\label{k3}
\kappa = 2-  \int_{0}^{\infty}dt \left[ \frac{\cosh\lr{t/4g}}{\cosh\lr{t/2g}}
\Big(\Gamma^{\BES}_{-}(t)-\Gamma^{\BES}_{+}(t)\Big)+\frac{\sinh\lr{t/4g}}{\cosh\lr{t/2g}}
\Big(\Gamma^{\BES}_{-}(t)+\Gamma^{\BES}_{+}(t)\Big) \right].
\ee
The integrals entering \re{e3} and \re{k3} look similar to that for $\epsilon_1(g)$,
Eq.~\re{epsilon1}, and their calculation goes along the same lines as before. Namely, we replace the
factors involving the ratios of hyperbolic functions on the right-hand side of \re{e3} and \re{k3}
by their Fourier integrals (see Eqs.~\re{B1} and \re{B11}), apply the relations \re{G-BES} to
evaluate the contribution from the region $0\le u\le 1$ and express the remaining integral over
$1\le u <\infty$ in terms of $\Gamma^{\BES}_\pm(t)$. Going through calculation we find that the
leading contribution to $\epsilon_3$ and $\kappa$ as $g\to\infty$ is proportional to the scale $m$,
Eqs.~\re{m=int} (see also \re{Mass}),
\be\label{eps3=m}
\epsilon_3= \frac{\pi^4}{96 \kappa^2} m + O\left(\e^{-3\pi g}\right) \,, \qquad \kappa=
\frac{\pi}{2} m + O\left(\e^{-3\pi g}\right)\,,
\ee
so that the relation \re{aa} takes the form $a=j/m+\ldots$. Finally, we combine together the
relations \re{e-expansion}, \re{eps1=m} and \re{eps3=m} and obtain small $j$ expansion of the
scaling function at strong coupling as
\be
\epsilon(g,j) = j(-1 + m) + \frac{\pi^2}{24m} j^3 +\ldots\,,
\ee
with $m$ given by \re{mass}. This relation agrees with the string theory prediction by Alday and
Maldacena~\cite{AM07}. It can be also written as
\be\label{e-scal}
\epsilon(g,j)+j = m^2 \left[\frac{j}{m}+ \frac{\pi^2}{24} \lr{\frac{j}{m}}^3 + \ldots\right],
\ee
wherefrom we expect that the expansion of the scaling function runs in powers of $j/m$, or
equivalently, $\epsilon_k \sim m^{2-k}$ at strong coupling.

\section{Scaling function and nonlinear $\rm O(6)$ sigma model}

For $j\sim m$, the small $j$ expansion employed  in the previous section is not applicable. In this
section, we will show that for  $j\ll g$ and $j/m={\rm fixed}$, the scaling function $\epsilon(g,j)$
coincides with the energy density of the ground state of the two-dimensional $\rm O(6)$ sigma model.

\subsection{Exact solution of the $\rm O(6)$ sigma model}

The exact solution for the ground state energy in the two-dimensional $\rm O(6)$ sigma-model was
constructed in Refs.~\cite{HMN90}. It can be summarized as  follows. The energy density
$\epsilon_{\rm O(6)}$ in the ground state and the particle density $\rho$ are given by
\be\label{ex1}
\epsilon_{\rm O(6)} = \frac{m}{2\pi}\int_{-B}^B d\theta\, \g(\theta) \cosh\theta\,,\qquad \rho
=\frac1{2\pi}\int_{-B}^B d\theta\, \g(\theta)\,,
\ee
where the differential rapidity distribution $\g(\theta)$ has the support on the interval $[-B,B]$
and satisfies the integral equation
\be\label{ex2}
\g(\theta) = \int_{-B}^B d\theta'\, K(\theta-\theta') \g(\theta') + m \cosh\theta \,.
\ee
Here the kernel $K(\theta)= \lr{ \ln S(\theta)}'/(2\pi i)$ is related to a logarithmic derivative of
the exact $S-$matrix of the $\rm O(6)$ model~\cite{ZZ78}
\begin{align}
K(\theta) = \frac{1}{4\pi^2}\left[\psi\left(1+\frac{i
\theta}{2\pi}\right)+\psi\left(1-\frac{i\theta}{2\pi}\right)-\psi\left(\frac{1}{2}-\frac{i
\theta}{2\pi}\right)-\psi\left(\frac{1}{2}+\frac{i
\theta}{2\pi}\right)+\frac{2\pi}{\cosh{\theta}}\right]\,,
\end{align}
where $\psi(x)=\lr{\ln\Gamma(x)}'$ is the Euler psi-function. Later in this section we will
encounter its Fourier transform
\be\label{K-Fourier}
K(\theta) =  \frac{2}{\pi^2} \int_{0}^{\infty}dt \,\cos{\left(2\theta t/ \pi \right)}
\frac{\e^{t}+1}{\e^{2t}+1}\,.
\ee
For $\rho/m\ll 1$, or equivalently $B\to 0$, the ground state energy of the model is given by
\be\label{e-o6}
\epsilon_{\rm O(6)} =   m \rho + \frac{\pi^2}{6m} \rho^3 + \ldots
\ee
and it coincides with the energy density of a dilute, nonrelativistic Fermi gas with the particle
density $\rho$~\cite{HMN90}.

Comparing \re{e-scal} with \re{e-o6} we observe that  the scaling function at strong coupling is
related at small $j\ll m$ to the ground state energy of the $\rm O(6)$ model as \footnote{We note
that the absence of $O(j^2)$ term in the expansion of the scaling function, $\epsilon_2(g)=0$, is
related to vanishing of $O(\rho^2)$ term in the energy of nonrelativistic Fermi gas.}
\be\label{map}
\epsilon_{\rm O(6)} = \frac{\epsilon(g,j)+j}{2}\,,\qquad \rho=\frac{j}2\,.
\ee
We will show in this section that the relation \re{map} holds for arbitrary $j$ in the scaling limit
$g\to\infty$ and $j/m={\rm fixed}$. More precisely, we will demonstrate that the FRS equation for
the scaling function $\epsilon(g,j)$ can be written in the form of \re{ex1} and \re{ex2} upon
identification
\be\label{rho}
\g(\theta) = \frac{8}{\pi}\int_{-\infty}^\infty dt \, \cos\lr{2\theta t/\pi} \,\gamma_{\rm h}(t)\,,
\ee
or conversely
\be\label{g-Fourier}
\gamma_{\rm h}(t) = \frac1{8\pi}\int_{-B}^B d\theta\, \cos\lr{2\theta t/\pi} \,\g(\theta)\,,
\ee
with $B=a \pi/2$.

\subsection{Rapidity distribution}

Let us first demonstrate that the Fourier transform of the function $\gamma_{\rm h}(t)$  fulfills
the same integral equation \re{ex2} as the rapidity distribution density for the $\rm O(6)$ model.

We recall that $\gamma_{\rm h}(t)$ satisfies the integral equation \re{gh-new} with the kernel
$K_{\rm h}(t,t')$ given by \re{K_h}. The  Fourier transform of this kernel can be easily evaluated
\be
\int_{-\infty}^\infty dt\, \e^{ikt} K_{\rm h}(t,t') = \frac12 \cos(kt') \theta(a^2-k^2)\,,
\ee
with the step function $\theta(x)$ equal to $1$ for $x\ge 0$ and $0$ otherwise. Making use of this
identity we perform Fourier transformation of both sides of \re{gh-new} and find that the function
$\g(\theta)$ introduced in \re{rho} vanishes for $\theta^2>B^2$ with $B=a\pi/2$. This property
should not be surprising since, in the Bethe Ansatz approach to the scaling
function~\cite{BGK06,FRS}, the function $\g(\theta)$ describes the distribution of holes%
\footnote{This function differs from the one introduced in \cite{FRS} by normalization,
$\chi(\theta)=2j \rho_{\rm h}(k)$ with $\theta=k \pi/2$.} which condense on the interval $[-B,B]$.
For $-B\le \theta\le B$ the function $\g(\theta)$ is given by
\be\label{rho=sum}
\frac{\pi}{8}\g\left(\theta\right)= \frac{1}{2} +I(\theta)- 2\int_{0}^{\infty}dt
\,\frac{\cos\lr{kt}}{\e^{t}-1} \ \left(\gamma_{\rm h}(t)-\frac{j}{8}\,\e^{{t}/{2}}
J_{0}(2gt)\right),
\ee
where $\theta=k\pi/2$ and the notation was introduced for
\be
I(\theta)=-\frac{g}2\int_{0}^{\infty}dt \, \frac{\cos(kt)}{\sinh(t/2)} \gamma(2gt) \,.
\ee
Let us rescale the integration variable, $t\to t/2g$, and apply \re{gG} to
eliminate $\gamma(t) = \gamma_+(t)+\gamma_-(t)$ in favor of  $\Gamma_\pm(t)$
\begin{align} \label{I-int}
I(\theta)
=&-\frac{1}{4}\int_{0}^{\infty}dt\, \cos \left({k t}/{2g}\right) \bigg[
\frac{\cosh\lr{t/4g}}{\cosh\lr{t/2g}}\Big(\Gamma_{-}(t)-\Gamma_{+}(t)\Big) +
\frac{\sinh\lr{t/4g}}{\cosh\lr{t/2g}}\Big(\Gamma_{-}(t)+\Gamma_{+}(t)\Big)\bigg].
\end{align}
We already encountered similar integral in Sect.~3.2.  The important difference is that the
functions $\Gamma_\pm(t)$ are now defined for $j\neq 0$. This does not affect however the general
scheme that we followed in Sect.~3.2 and  calculation goes along the same lines as before (see
Appendix B for details). It leads to the following relation in the scaling limit $g\to\infty$ and
$j/m={\rm fixed}$ (with $\theta=k\pi/2$)
\be\label{A+B+C}
I(\theta)=- \frac12 + \frac{\pi}{8}m \cosh\theta+ 4 \int_{0}^{\infty}dt \, \frac{
\cos(kt)\e^{{t}/{2}}}{\e^{t}+\e^{-t}}  \, \frac{1}{\e^{t}-1} \left(\e^{{t}/{2}} \gamma_{\rm h}(t) -
\frac{j}{8} \ J_{0}(2gt)\right)\,.
\ee
It is important to emphasize that this relation was derived in Appendix B under assumption that
\be\label{asumption}
B < \pi g\,.
\ee
Taking into account the identity $\theta=k\pi/2$, we find that this condition can be reformulated as
a requirement that the interval $[-a,a]$  on the real $k-$axis in which the holes condense should be
located inside the gap $[-2g,2g]$ corresponding to the magnon density at strong coupling described
by the BES equation.

Substitution of \re{A+B+C} into \re{rho=sum} yields
\be\label{xi}
\g(\theta) =  m\cosh\theta+ \frac{16}{\pi} \int_{0}^{\infty}dt \,\cos(kt) \frac{1+\e^{t}}{1+\e^{2t}}
\gamma_{\rm h}(t) + \frac{2j}{\pi} \int_{0}^{\infty}dt\,\cos(kt) \frac{\sinh\lr{t/2}}{\cosh {t}} \,
J_{0}(2gt)\,.
\ee
The last term on the right-hand side of this relation is subleading in the scaling limit (see
Eqs.~\re{Bnl} and \re{Bl}) and can be neglected. Then, we replace $ \gamma_{\rm h}(t)$ by its
Fourier integral \re{g-Fourier},
\be
\g(\theta) =  m\cosh\theta+ \frac{2}{\pi^2}\int_{-B}^B d\theta'\,\g(\theta') \int_{0}^{\infty}dt
\,\cos(2\theta t/\pi) \cos\lr{2\theta' t/\pi} \frac{1+\e^{t}}{1+\e^{2t}}\,,
\ee
integrate over $t$ with a help of \re{K-Fourier} to find that $\g(\theta)=\g(-\theta)$ satisfies the
same integral equation \re{ex2} as the rapidity distribution in the  $\rm O(6)$ model!

Finally, it follows from \re{ex1} and \re{g-Fourier} that the density of particles in the $\rm O(6)$
model is related to the scaling parameter $j$ as
\be
\rho = \frac1{2\pi} \int_{-B}^B d\theta\, \g(\theta) = 4\gamma_{\rm h}(0) = \frac{j}2\,,
\ee
where in the last relation we applied \re{gh-even}.

\subsection{Energy of the ground state}

It remains to show that the scaling function $\epsilon(g,j)$ is related to the energy of the ground
state of the $\rm O(6)$ model \re{ex1}  through relation \re{map}.

As follows from  \re{epsilon-g}, the scaling function admits the following representation
\be\label{epsilon-g1}
\epsilon(g,j)   =
 16{g}  \int_{0}^{\infty}\frac{dt}{t} \
\frac{\gamma_+^{\BES}(2gt)-\gamma_-^{\BES}(2gt)}{\sinh(t/2)} \left(
  \gamma_{\rm h}(t) -\gamma_{\rm h}(0) \right)
 +j \epsilon_1(g) \,,
\ee
where we separated terms linear in $j$ into the function $\epsilon_1(g)$ given by \re{eps13}. As
before, we eliminate $\gamma^{\BES}_\pm(t)$ in favor of $\Gamma^{\BES}_\pm(t)$ with a help of
\re{gG} and replace $\gamma_{\rm h}(t)$ by the Fourier integral \re{g-Fourier}
\be\label{e=int}
\epsilon(g,j) = \frac{2g}{\pi}\int_{-B}^Bd\theta\,\g(\theta)\, E(\theta) + j \,\epsilon_1(g)\,,
\ee
where the notation was introduced for (with $\theta=k\pi/2$)
\begin{align}
E(\theta) = \int_0^\infty\frac{dt}{t}\big({1-\cos(kt/2g)}\big) \bigg[\ &
\frac{\cosh\lr{t/4g}}{\cosh\lr{t/2g}}\Big(\Gamma^{\BES}_{-}(t)+\Gamma^{\BES}_{+}(t)\Big)
\\ \notag
-& \frac{\sinh\lr{t/4g}}{\cosh\lr{t/2g}}
\Big(\Gamma^{\BES}_{-}(t)-\Gamma^{\BES}_{+}(t)\Big)\bigg]\,.
\end{align}
We observe similarity of this integral with the one entering \re{I-int}. Going through the same
steps as before, we get at large $g$,
\be
E(\theta)  = \frac{m}{2g}\lr{\cosh\theta-1}\,,
\ee
with $m$ being the mass gap \re{mass}. Plugging this expression into \re{e=int} and making use of
\re{eps1=m}, we evaluate the scaling function as
\be
\epsilon(g,j) = \frac{m}{\pi}\int_{-B}^Bd\theta\,\g(\theta)\,\lr{\cosh\theta-1} + j (-1+m) =
\frac{m}{\pi}\int_{-B}^Bd\theta\,\g(\theta)\, \cosh\theta -j\,,
\ee
in a perfect agreement with \re{map} and \re{ex1}.

Thus, we demonstrated that, in the limit $g\to\infty$ with $j/m=\rm fixed$, the scaling function
$\epsilon(g,j)$ is related to the energy density in the ground state of the two-dimensional $\rm
O(6)$ sigma-model \re{map} and, therefore, it can be found from the exact solution of this model
constructed in \cite{HMN90}. We should keep in mind however that this result was based on the
assumption \re{asumption} which need to be checked. Let us first consider the region $j\ll m$ in
which case $a=j/m +\ldots$ and, therefore, $B=a\pi/2$ automatically satisfies \re{asumption}. The
relation \re{asumption} becomes nontrivial for $j\gg m$. In this region, the exact solution of the
${\rm O(6)}$ model leads to~\cite{HMN90}
\be
\epsilon_{\rm O(6)} = \frac{j^2}{2}  \left[ \frac{\pi}{8\ln(j/m)}+
O\lr{\frac{\ln\ln(j/m)}{\ln^2(j/m)}}\right].
\ee
This expression is valid for $j \gg m$ independently on whether $j\gg g$ or $j\ll g$ whereas the
string theory consideration~\cite{AM07} suggests that the relation between $\epsilon(g,j)$ and
$\epsilon_{\rm O(6)}$ should only hold for $j\ll g$. Remarkably enough, the relation \re{asumption}
leads to the condition for $j$ on the gauge theory side compatible with $j\ll g$. Indeed, for $j\gg
m$ we expect from the exact solution of the ${\rm O(6)}$ model~\cite{PW83,FR85,HMN90} that the
$B-$parameter scales as $B\sim \ln (j/m)-c \ln\ln(j/m)$ (with constant $c$ that need to be calculated)  leading to
\be
 j \sim m g^{c} \e^{B} \sim g^{c+1/4}\e^{B-\pi g}\,,
\ee
where in the second relation we used the expression for the mass gap \re{mass}. We conclude that the
condition \re{asumption} implies that $j$ is exponentially small in $g$. We expect that the constant  $c$ should be equal to $c=3/4$ leading to the condition $j/g\ll 1$, in agreement with string theory prediction.

\section{Conclusions}

In this paper, we have studied  anomalous dimensions of high-twist Wilson operators in a particular
limit \cite{BGK06} when  their Lorentz spin grows exponentially with the twist. The anomalous
dimensions have a nontrivial scaling behavior in this limit and conjectured integrability of the
dilatation operator in  $\mathcal{N}=4$ SYM theory leads to the FRS equation \cite{FRS} for the
corresponding scaling function $\epsilon(g,j)$. It is believed that the function $\epsilon(g,j)$
defined as a solution to this equation should interpolate between the weak coupling result based on
explicit perturbative calculation in $\mathcal{N}=4$ SYM and the strong coupling result predicted
within the gauge/string duality.

We solved the FRS equation at strong coupling in the scaling limit, $g\to\infty$ and $j/m={\rm
fixed}$, and found that, quite remarkably, the anomalous dimensions of high-twist operators in
four-dimensional conformal invariant $\mathcal{N}=4$ SYM theory  are described by two-dimensional
nonlinear $\rm O(6)$ model which is asymptotically free at short distances and develops a mass gap
in the infrared. This result is in a perfect agreement with the proposal  by Alday and Maldacena
\cite{AM07} who argued, based on the dual string consideration, that the scaling function
$\epsilon(g,j)$ should coincide with the energy density of the $\rm O(6)$ model embedded into the
$\rm AdS_5\times S^5$ model.

In the Bethe Ansatz approach, the scaling function is determined by two sets of rapidities
describing magnons and holes~\cite{BGK06}. In the scaling limit, the rapidities condense on the real
axis and their distribution densities satisfy the system of coupled integral equations \cite{FRS}.
We found that solutions to these equations at strong coupling depend on a `hidden' nonperturbative
scale $m$ which determines the mass spectrum of hole excitations. The same scale has been previously
identified in the strong coupling expansion of the cusp anomalous dimension \cite{BKK07}. We
demonstrated by explicit calculation that, firstly, the scale $m$ coincides with the exact mass gap
of the two-dimensional $\rm O(6)$ sigma-model and, secondly, the  dynamics of holes at strong
coupling is described by the thermodynamical Bethe Ansatz equations for the same model.

At weak coupling, $\epsilon(g,j)$ is a bi-analytical function of $g$ and $j$ \cite{FRS}. At strong
coupling, due to appearance of the mass gap,  $\epsilon(g,j)$ has a nontrivial `phase diagram' in
the $(g,j)-$plane. Namely, for $m\ll g$, the function $\epsilon(g,j)$  has different behavior for
$j\ll m\ll g$ and $m\ll j\ll g$~\cite{AM07}. In both regimes, $\epsilon(g,j)$ coincides with the
energy density in the ground state of  $\rm O(6)$ model: For $j\ll m\ll g$, the function
$\epsilon(g,j)$ admits an expansion in powers of $j/m$ which reflects nontrivial properties of the
$\rm O(6)$ model in the infrared. For $m\ll j\ll g$, in accordance with the asymptotic freedom in
this model, $\epsilon(g,j)$ is given by the perturbative series in $1/g$ with coefficients
proportional to $j^2$ and enhanced by powers of $\ln(j/g)$.

We would like to stress that the FRS equation predicts the scaling function $\epsilon(g,j)$ for
arbitrary $g$ and $j$.  The $\rm O(6)$ sigma-model only describes the leading asymptotic behavior of
$\epsilon(g,j)$  in the limit $g\to\infty$ and $j\ll g$. It would be interesting to
investigate whether the relation with the $\rm O(6)$ model also exists at the level of subleading
corrections. Based on our experience with the cusp anomalous
dimension~\cite{BES,Benna06,Alday07,Kostov07,BKK07,B08,KSV08}, we expect that the transition from
strong to weak coupling regime in  $\epsilon(g,j)$ should occur for $g\sim 1$. In this region, the
hierarchy between the scales $m$ and $g$ disappears and the relation to the $\rm O(6)$ sigma-model
is lost. The crossover between strong and weak couplings deserves additional study. Finally, the
scaling function $\epsilon(g,j)$ has a nontrivial behavior \cite{BGK06} at strong coupling in the
region  $j
> g$ (with $N,L\to\infty$), which is left beyond the scope of the present paper. As was already
mentioned in the Introduction, the anomalous dimension in this region does not grow logarithmically
with $N$ anymore but has instead  a BMN like behavior \re{BMN}. It would be interesting to reproduce
this behavior from the FRS equation.

\section*{Acknowledgements}

We would like to thank Marcus Benna, Igor Klebanov, Juan Maldacena and Mikhail Shifman for
interesting discussions. B.B. thanks the Institute for Nuclear Theory at the University of
Washington for its hospitality and the Department of Energy for partial support in the final stage
of the work. This research was supported in part by the French Agence Nationale de la Recherche
under grant ANR-06-BLAN-0142.

\appendix
\setcounter{section}{0}
\setcounter{equation}{0}
\renewcommand{\theequation}{A.\arabic{equation}}

\section*{Appendix A:\ \ Derivation of the scaling function}

This appendix contains a derivation of the relation \re{epsilon-g}.
According to \re{ff} and \re{f=g1}, the scaling function $\epsilon(g,j)$ is given by
\be\label{A1}
\epsilon(g,j) = f(g,j)-f(g,0)=16g^2 \left[\gamma_1(g,j)-\gamma_1(g,0)\right],
\ee
where $\gamma_1(g,j)$ enters into the Bessel series expansion of function $\gamma_-(t;j)$,
Eq.~\re{Bessel}. The function  $\gamma_-(t;j)$ satisfies the  integral equations \re{FRS2} and
satisfies Wronskian-like relations.

To derive these relations we choose some reference $j'$, multiply both sides of the two relations in
\re{FRS2} by the coefficients $(2n-1)\gamma_{2n-1}(g,j')$ and $(2n) \gamma_{2n}(g,j')$,
respectively, and sum over $n\ge 1$. Then, we convert the sums into the functions $\gamma_\pm(t;j')$
using the definition \re{Bessel} and subtract the second relation from the first one to obtain
(after rescaling $t\to t/2g$)
\begin{align}
\int_0^\infty\frac{dt}{t}\bigg[\frac{\gamma_-(t;j)\gamma_-(t;j')-\gamma_+(t;j)\gamma_+(t;j')}{1-\e^{-t/2g}}
+ \frac{\gamma_-(t;j)\gamma_+(t;j')+\gamma_+(t;j)\gamma_-(t;j')}{\e^{t/2g}-1}\bigg]
\\ \notag
=  \gamma_1(g,j') +2\sum_{n\ge 1} \big[(2n-1)
h_{2n-1}(g,j)\gamma_{2n-1}(g,j')-(2n)h_{2n}(g,j)\gamma_{2n}(g,j')\big].
\end{align}
The expression on the left-hand side  is invariant under exchange $j \leftrightarrow j'$ and the
same should be true on the right-hand side. Then, we replace $h_n$ by their definition \re{h_n} and
get
\be\label{diff-gamma}
\gamma_1(g,j)-\gamma_1(g,j') = \frac2{g}  \int_{0}^{\infty}\frac{dt}{t} \
\frac{\gamma_+(2gt;j')-\gamma_-(2gt;j')}{\e^{t}-1} \left[
  \e^{{t}/{2}}  \gamma_{\rm h}(t;j) -\frac{j}{8}J_0(2gt)  \right]- \lr{j\leftrightarrow j'}\,,
\ee
where we indicated explicitly the dependence of the $\gamma-$functions on the scaling parameters. We
note that for $j'=0$ the relation \re{diff-gamma} can be used to evaluate the scaling function
\re{A1}. We take into account that $\gamma_{\rm h}(t;j'=0)=h_n(g,j'=0)=0$ and $\gamma_\pm(t;j'=0)
\equiv \gamma_\pm^{\BES} (t)$ to obtain
 \begin{align}\label{A2}
 \epsilon(g,j)   =
 32{g}  \int_{0}^{\infty}\frac{dt}{t} \
\frac{\gamma_+^{\BES}(2gt)-\gamma_-^{\BES}(2gt)}{\e^{t}-1} \left[
 \lr{ \e^{{t}/{2}}  \gamma_{\rm h}(t;j)-\frac{j}8} +\frac{j}{8}\big({1-J_0(2gt)}\big)  \right].
 \end{align}
Here the integral involving the Bessel function can be further simplified by making use of the
identity $1-J_0(z)=2\sum_{n\ge 1} J_{2n}(z)$ leading to
\begin{align}\label{A3}
& 8gj\sum_{n\ge 1} \int_0^\infty \frac{dt}{t}
\frac{\gamma_+^{\BES}(2gt)-\gamma_-^{\BES}(2gt)}{\e^{t}-1}
 J_{2n}(2gt)
 \\ \notag
& \qqqquad\qqquad =-4gj \int_0^\infty \frac{dt}{t} \gamma_+^{\BES}(2gt)   \lr{1-J_0(2gt)}= -4gj
\int_0^\infty \frac{dt}{t} \gamma_+^{\BES}(2gt) \,,
\end{align}
where in the first relation we applied the integral equations \re{FRS2} for $j=0$ and in the second
relation used Bessel series representation \re{Bessel} together with the orthogonality condition
\re{ortho}. Combining together the relations \re{A2} and \re{A3} we arrive at \re{epsilon-g}.

\renewcommand{\theequation}{B.\arabic{equation}}
\setcounter{section}{0} \setcounter{equation}{0}

\section*{Appendix B:\ \ Strong coupling expansion}

In this Appendix we present some details of the strong coupling expansion of the scaling function.
In our analysis we encountered integrals like Eqs.~\re{epsilon1} and \re{I-int} which involve the
product of ratios of hyperbolic functions and linear combinations of $\Gamma_\pm(t)$. To begin with,
we replace the former by their Fourier integrals with a help of identities (with $g>0$)
\begin{align}\label{B1}
\cos(kt) \frac{\cosh(t/4g)}{\cosh(t/2g)}  &= \sqrt{2} g \int_{-\infty}^\infty du\,
\cos(ut)\frac{\cosh(g\pi (u+k))}{\cosh(2g\pi (u+k))}\,,
\\ \notag
\cos(kt) \frac{\sinh(t/4g)}{\cosh(t/2g)}  &= \sqrt{2} g \int_{-\infty}^\infty du\,
\sin(ut)\frac{\sinh(g\pi (u+k))}{\cosh(2g\pi (u+k))}\,,
\end{align}
valid for arbitrary real $k$ and $t$. To verify these relations, one closes the integration contour
into the lower or upper half-plane depending on the sign of $t$ and picks up residues coming from
the denominator. The calculation of $\epsilon_3$ also involves the integrals
\begin{align}\label{B11}
t \frac{\cosh(t/4g)}{\cosh(t/2g)}  &= -\sqrt{2} g \int_{-\infty}^\infty du\,
\sin(ut)\frac{d}{du}\lr{\frac{\cosh(g\pi u)}{\cosh(2g\pi u)}}\,,
\\ \notag
t \frac{\sinh(t/4g)}{\cosh(t/2g)}  &= \sqrt{2} g \int_{-\infty}^\infty du\,
\cos(ut)\frac{d}{du}\lr{\frac{\sinh(g\pi u)}{\cosh(2g\pi u)}}\,.
\end{align}
In what follows we only calculate in detail $\epsilon_1$ since the analysis of $\epsilon_3$ is
essentially the same. Applying \re{B1} for $k=0$ we find from \re{epsilon1}
\begin{align}\label{B2}
\epsilon_{1}(g) =  - {2\sqrt{2}g^2} \bigg\{  & \int_{-\infty}^{\infty}du \ \frac{\cosh\lr{g\pi
u}}{\cosh\lr{2g\pi u}} \int_{0}^{\infty}\frac{dt}{t} \ \Big(1-\cos\lr{ut}\Big)
\Big[\Gamma^{\BES}_{-}(t)+\Gamma^{\BES}_{+}(t)\Big]
 \\ \notag
+& \int_{-\infty}^{\infty}du \ \frac{\sinh\lr{g\pi u}}{\cosh\lr{2g\pi u}}
\int_{0}^{\infty}\frac{dt}{t} \ \sin\lr{ut} \
\Big[\Gamma^{\BES}_{-}(t)-\Gamma^{\BES}_{+}(t)\Big]\bigg\}\,.
\end{align}
Let us split the $u-$integrals into sum of two terms corresponding to $u^2\le 1$ and $u^2>1$. In the
first one, we replace $\lr{1-\cos(ut)}/t = \int_0^u dv\,\sin(vt)$ and $\sin(ut)/t  = \int_0^u dv\,
\cos(vt)$ and, then, evaluate the $t-$integral with a help of \re{G-BES}. The resulting expression
for $\epsilon_1(g)$ is given by \re{eps1=3terms}. To find the large $g$ expansion of the first term
on the right-hand side of \re{eps1=3terms} we just do an opposite, that is rewrite the integral over
$-1\le u\le 1$ as a difference of two integrals over $-\infty <u < \infty$ and $u^2>1$
\begin{align} \notag
\int_{-1}^{1}du \, u \frac{\sinh\lr{g\pi u}}{\cosh\lr{2g\pi u}} & = \int_{-\infty}^{\infty}du \, u
\frac{\sinh\lr{g\pi u}}{\cosh\lr{2g\pi u}} -2 \int_{1}^{\infty}du \, u \frac{\sinh\lr{g\pi
u}}{\cosh\lr{2g\pi u}}
\\
&=\frac1{4g^2\sqrt{2}} - 2\frac{1+\pi g }{(\pi g)^2}  \e^{-g\pi} +O(\e^{-3g\pi }) \,.
\end{align}
In the remaining two terms in  \re{eps1=3terms}, we replace hyperbolic functions by their leading
large $g$ asymptotics, integrate over $u$ by parts, take into account \re{G-BES} to evaluate the
boundary term and arrive at \re{eps1=m} with $m$ given by
\begin{align}\label{Mass}
m = \frac{2^{7/2} g}{\pi} \bigg\{\frac{\e^{-\pi g}}{\pi g}- & \frac12 \int_{1}^{\infty}du\, \e^{-\pi
g u} \int_{0}^{\infty}dt
\\ \notag
&\times \bigg[ \cos(ut) \Big(\Gamma^{\BES}_{-}(t)-\Gamma^{\BES}_{+}(t)\Big) + \sin(ut)
\Big(\Gamma^{\BES}_{-}(t)+\Gamma^{\BES}_{+}(t)\Big) \bigg]\bigg\}.
\end{align}
Integration over $u$ leads to \re{m=int}. Then, we change the integration variable in \re{m=int} as
$t=g\pi \tau$ and take into account \re{G-as} to get
\be\label{B10}
m=\frac{2^{7/2}}{\pi^2} \e^{-\pi g}+  \Delta m\,,
\ee
with $\Delta m$ given by a complicated twofold integral
\be
\Delta m  =  g \e^{-\pi g}\frac{2^{5/2} }{\pi^2}\int_0^\infty \frac{d\tau\, \e^{-i\pi/4}}{\tau+i}
\frac{ \Gamma(3/4)\Gamma(1+i\tau/4)}{\Gamma(3/4+i\tau/4)} \int_{-1}^{1}du \,
\left(\frac{1+u}{1-u}\right)^{1/4} \e^{ i\pi g (1-u) \tau}+ \text{c.c.}
\ee
To find its large $g$ asymptotics, we use Mellin-Barnes representation for $\e^{i\pi g (1-u) \tau}$ and
integrate over $u$ to obtain
\be
\Delta m= g \e^{-\pi g}\frac{4}{\pi} \int_{-\delta-i\infty}^{-\delta+i\infty}\frac{dj}{2\pi i}(2\pi
g)^{j} \frac{\Gamma(-j) \Gamma(j+3/4)}{ \Gamma(j+2)} \int_0^\infty \frac{d\tau\,
(-i\tau)^{j}}{\tau+i}  \frac{\e^{-i\pi/4}\Gamma(1+i\tau/4)}{\Gamma(3/4+i\tau/4)} + \text{c.c.}
\ee
At large $g$ we deform the integration contour to the left and pick up the contribution of poles  at
negative $j$. The first two poles closest to the origin are located at $j=-3/4$ and $j=-1$ with the
second one coming from integration at small $\tau$. Going through calculation of residues we get
\be
\Delta m =  g \e^{-\pi g}\frac{4}{\pi} \left[ \lr{2\pi
g}^{-3/4}\frac{2^{-1/2}\pi^{2}}{\Gamma(5/4)}-\lr{2\pi g}^{-1} {2^{5/2}}   +
O\left(g^{-7/4}\right)\right]\,.
\ee
Substitution of this relation into \re{B10} yields the desired expression for the mass gap
\re{mass}.

The evaluation of the integral $I(\theta)$ defined in \re{I-int} goes as follows.
We replace the factors in front of $\Gamma_\pm(t)$ functions on the right-hand side of
\re{I-int} by their Fourier integrals (see Eq.~\re{B1}), split the $u-$integral into sum of two
terms corresponding to $u^2\le1$ (A+B) and $u^2>1$ (C) and, then, simplify the first term using
\re{G-int}. In this way, we find
\be\label{I}
I(\theta) = I_{\rm A}(\theta) +  I_{\rm B}(\theta)+  I_{\rm C}(\theta)\,,
\ee
where
\begin{align}
I_{\rm A}(\theta)  &=   - \frac{g}{\sqrt{2}}\int_{-1}^{1}du \, \frac{\cosh\lr{g\pi
u+\theta}}{\cosh\lr{2g\pi u+2\theta}}\,,
\\[4mm] \notag
I_{\rm B}(\theta)  &= 2\sqrt{2}g \int_{-1}^{1}du\int_{0}^{\infty}\frac{dt}{\e^{t}-1}
\left(\e^{{t}/2} \gamma_{\rm h}(t) - \frac{j}{8} \ J_{0}(2gt)\right)
\\ \notag
& \hspace*{32mm} \times \left[ \frac{\cosh\lr{g\pi u+\theta}}{\cosh\lr{2g\pi u+2\theta}}
\cos\lr{2gut} +  \frac{\sinh\lr{g\pi u+\theta}}{\cosh\lr{2g\pi u+2\theta}} \sin\lr{2gut}\right]\,,
\end{align}
\begin{align}\notag
I_{\rm C}(\theta) = -\frac{g}{2\sqrt{2}} \int_{1}^{\infty}du\int_{0}^{\infty}dt\, \bigg[\ &
\frac{\cosh\lr{g\pi u+\theta}}{\cosh\lr{2g\pi u+2\theta}} \cos({ut}) \
\Big(\Gamma_{-}(t)-\Gamma_{+}(t)\Big)
\\ \notag
+&  \frac{\sinh\lr{g\pi u+\theta}}{\cosh\lr{2g\pi u+2\theta}}  \sin({ut}) \
\Big(\Gamma_{-}(t)+\Gamma_{+}(t)\Big)\bigg]  + (\theta \to -\theta).
\end{align}
Here $I_{\rm A}(\theta)$ and $I_{\rm B}(\theta)$ originate from the first and the second terms on
the right-hand side of \re{G-int}, respectively.

To find asymptotic behavior of $I_{\rm A}(\theta)$ at large $g$ we split the $u-$integral into the
difference of two integrals over $-\infty < u <\infty$ and $u^2\ge 1$. The former integral can be
easily done with a help of identity \re{B1}. Calculating the second integral as well as $I_{\rm
C}(\theta)$, we replace the ratios of hyperbolic functions by their leading asymptotic behavior at
large $g$ assuming that $(g\pi u \pm\theta)$ does not change sign inside the integration region.
This assumption is justified provided that $-\pi g < \theta < \pi g$, or equivalently $-2g < k <
2g$. Since the integral \re{I} is defined on the support interval $-B \le \theta \le B$ of the hole
density $\chi(\theta)$,  the above mentioned condition is automatically satisfied for $B<\pi g$, or
equivalently $a<2g$.  Assuming that this condition is fulfilled, we find after some algebra
\be\label{A+C}
I_{\rm A}(\theta)+I_{\rm C}(\theta) = - \frac12 + \frac{\pi}{8}m(j)\cosh\theta\,,
\ee
where the parameter $m(j)$ is given by
\begin{align}\label{mj}
m(j) & = \frac{8\sqrt{2}}{\pi^2} \e^{-\pi g}-\frac{8g}{\pi \sqrt{2}} \int_{1}^{\infty}du\, \e^{-\pi
g u} \int_{0}^{\infty}dt
\\ \notag
& \hspace*{30mm} \times\left[\cos\lr{ut} \ \Big(\Gamma_{-}(t)-\Gamma_{+}(t)\Big) + \sin\lr{ut} \
\Big(\Gamma_{-}(t)+\Gamma_{+}(t)\Big) \right].
\end{align}
Here the dependence on $j$ resides inside the functions $\Gamma_\pm(t)$ only.  In the previous
section, we encountered the same expression for $j=0$ and identified $m(0)=m$ as the mass gap of the
$\rm O(6)$ sigma-model, Eq.~\re{mass}. We notice that the $u-$integral on the right-hand side of
\re{mj} scales as $\e^{-\pi g}$ at large $g$. This allows us to neglect in the $t-$integral
corrections proportional to $j$ in the scaling limit $g\to\infty$ and $j/m={\rm fixed}$, that is to
substitute the functions $\Gamma_\pm(t)$ by their values at $j=0$,  leading to $m(j)= m(0)+\ldots$.
From the point of view of the integral equation \re{G-int} this amounts to neglecting the second
term on the right-hand side of \re{G-int} as being proportional to $j$.

To evaluate $I_{\rm B}(\theta)$, we rewrite the $u-$integral as the difference of two integrals over
$-\infty < u < \infty$ and $u^2\ge 1$. The former integral can be easily evaluated with a help of
identity \re{B1} while the latter one produces the contribution subleading in the scaling limit
provided that $-\pi g < \theta < \pi g$ and, therefore, can be safely neglected leading to
\be\label{B}
I_{\rm B}(\theta) = 4 \int_{0}^{\infty}dt \, \frac{ \cos(2\theta t/\pi)\e^{{t}/{2}}}{\e^{t}+\e^{-t}}
\, \frac{1}{\e^{t}-1} \left(\e^{{t}/{2}} \gamma_{\rm h}(t) - \frac{j}{8} \ J_{0}(2gt)\right)\,.
\ee
Combining together the relations \re{I}, \re{A+C} and \re{B} we arrive at \re{A+B+C}.

Calculating \re{xi} we encountered the following integral
\begin{align}\notag
 \frac{2j}{\pi} \int_{0}^{\infty}dt\,\cos(2\theta t/\pi) \frac{\sinh\lr{t/2}}{\cosh {t}} \,
J_{0}(2gt) & =\frac{\sqrt{2}j}{\pi}\int_{-\infty}^\infty du\frac{\sinh(\pi g u+\theta)}{\cosh(2\pi g
u+2\theta)}\int_0^\infty dt\, J_0(t)\sin(ut)
\\ \label{Bnl}
&=\frac{\sqrt{2}j}{\pi}\int_{-\infty}^\infty du\frac{\sinh(\pi g u+\theta)}{\cosh(2\pi g
u+2\theta)}\frac{\theta(u^2-1)}{\sqrt{u^2-1}}\,,
\end{align}
where we rescaled the integration variable $t\to t/2g$, applied the second relation in \re{B1} and
used the properties of Bessel functions. Expanding hyperbolic functions at large $g$ we evaluate the
integral as
\be\label{Bl}
\frac{2\sqrt{2}j}{\pi}\cosh\theta \int_1^\infty \frac{du\,\e^{-\pi g u}}{\sqrt{u^2-1}} =
\frac{2}{\pi}   {j\,g^{-1/2} \e^{-\pi g}} \cosh\theta + \ldots
\ee
In the limit $g\to\infty$ and $j/m={\rm fixed}$, it scales as $m^2$ and, therefore, provides a
subleading contribution to the right-hand side of \re{xi}.

\end{document}